# Transformation of perturbative series into complex phases and elimination of secular divergences in time-dependent perturbation theory in quantum mechanics

Q. H. Liu*

School for Theoretical Physics, School of Physics and Electronics, Hunan University,

Changsha 410082, China

(Dated: July 8, 2016)

**Abstract**

The difficulty that the probabilities infinitely increase with time as time is long enough in time-dependent perturbation theory for some quantum systems is resolved by means of simply transforming the perturbative series into natural exponential functions of the re-summed perturbative series. Three exactly solvable models are taken to check our new formulation, and excellent agreements with the exact solution are achieved.

# I. Introductions

Textbooks of quantum mechanics on the time-dependent perturbation theory (TDPT) seldom treat the secular divergences seriously, but simply claim that validity of naïve TDPT is usually limited to short periods of time. [1,2] For instance, for some physical processes the perturbation persists for a long period of time, Pauli comments that "we have to avoiding the perturbation method" [1] to deal with them. For the spin resonance with weak perturbation of driven frequency $\omega_1$ and the perturbation acts for time interval $T$ that can be arbitrarily large, where the TDPT predicts a divergent consequence, Ballentine comments on this problem: "It is apparent that perturbation theory will be accurate at resonance only if $|\omega_1 T| \ll 1$. No matter how weak the perturbing field may be, perturbation theory will fail if the perturbation acts for a sufficiently long time." [2] All of these observations are true but nevertheless superficial, for they are associated with the naïve form of the TDPT. [3,4]

I will show in this paper that, the perturbative series (PS) in powers of perturbative strength in the naïve form of the TDPT in quantum mechanics can be converted into the natural exponential functions of another PS also in powers of the perturbative strength. Once it is done, there is no secular divergence. In fact, in order to remove the these divergences, there were attempts to revisit the fundamentals of the TDPT, [3,4] or to utilize the modern technique such as renormalization group. [5] However, all these developments appear too advanced to be an essential part of the lecture notes on elementary quantum mechanics.

For clearly illustrating what secular divergences associated with the perturbation theory are and how I eliminate them, let us consider a toy system, the following first-order differential equation of $x(t)$ with respect to $t$,

$$\begin{cases} i\dfrac{dx(t)}{dt} = (\omega+\varepsilon)x(t) \\ x(0)=1 \end{cases},\quad (t\geq 0) \tag{1}$$

where $\omega$ and $\varepsilon$ are two parts of the frequency and $|\omega|\gg|\varepsilon|$. On one hand, the exact solution to this equation is given by $x(t)=e^{-i(\omega+\varepsilon)t}$. Once the phase factor part $e^{-i\varepsilon t}$ is expanded in powers of $\varepsilon$, the approximate solution is,

$$x(t)\approx e^{-i\omega t}\left(1-i\varepsilon t-\varepsilon^{2}t^{2}+...\right). \tag{2}$$

On other hand, by solving equation (1) with the straightforward application of the perturbation theory, we mean to seek solution of following PS expansion in powers of parameter $\varepsilon$,

$$x(t)=x^{(0)}(t)+\varepsilon x^{(1)}(t)+\varepsilon^{2}x^{(2)}(t)+...=1+\varepsilon x^{(1)}(t)+\varepsilon^{2}x^{(2)}(t)+.... \tag{3}$$

We have another approximate solution $x_{a}(t)$ of $x(t)$, accurate up to the first order of $\varepsilon$,

$$x_{a}(t)\approx e^{-i\omega t}(1-i\varepsilon t). \tag{4}$$

The secular divergence exhibits in itself in comparison of the absolute value of this approximate solution with that of the exact one,

$$|x(t)|=1,\quad |x_{a}(t)|\approx\sqrt{1+(\varepsilon t)^{2}}\xrightarrow{t\to\infty}|\varepsilon t|. \tag{5}$$

To be precise, the secular divergence means a linear dependence of $|x_{a}(t)|$ with time when the time is large. Because in general this divergence does not exist in physics, one is usually to conclude that the TDPT is only applicable for short periods of time such that $|\varepsilon t|\ll 1$.

Clearly, this divergence is purely mathematical, and can be easily removed by either using the natural exponential function of a series in powers of $\varepsilon$

$$x(t) = \exp\left(\varepsilon X^{(1)}(t) + \varepsilon^2 X^{(2)}(t) + ...\right), \tag{6}$$

where $\varepsilon X^{(1)}(t) + \varepsilon^2 X^{(2)}(t) + ... \left(= \ln\left\{1 + \varepsilon x^{(1)}(t) + \varepsilon^2 x^{(2)}(t) + ...\right\}\right)$ is in fact a re-summed series of the original PS, or seek more and more terms in the original PS. Explicitly, for the former, we have $1 - i\varepsilon t \approx e^{-i\varepsilon t}$, whereas for the latter, we have $\sum_r (-i\varepsilon t)^r / r! = e^{-i\varepsilon t}$. Both give almost the same result that is free from the secular divergences. Moreover, the former amounts to a renormalization of the frequency from $\omega$ into $\omega + \varepsilon$.

In the rest part of the paper, I will transform the PS into the complex phases to eliminate the secular divergences associated with the straightforward use of the TDPT in quantum mechanics. This paper is organized as what follows. In section II, the routine way of applying the TDPT to the quantum systems is outlined and a transformation of the PS into complex phases is presented. In section III, three exactly solvable models are taken to compare the approximate solutions that suffer from secular divergences with our solutions that are excellently agreement with the exact ones. In section IV a brief conclusion and discussion is given.

**II. An outline of the TDPT and transformation of the PS into complex phases**

**II. 1 An outline of the TDPT**

The standard formulation of the TDPT in quantum mechanics is available in every textbook on quantum mechanics, and an outline is given in the following.

The Hamiltonian $H$ can be split into two parts,

$$H = H_0 + h(t), \quad h(t) \neq 0, \text{ for } t > t_0, \tag{7}$$

where $H_0$ is the main part that does not contain time explicitly, and the time-dependent part $h(t)$ is a perturbation of the system, switching-on at $t_0$ which is usually set to be zero. The energy eigenkets $|n\rangle$ and the energy eigenvalues $E_n$, determined by following eigenvalue function,

$$H_0|n\rangle = E_n|n\rangle, \tag{8}$$

are completely known. The eigenket set $\{|n\rangle\}$ spans a Hilbert space. At initial time $t_0 = 0$, any state ket $|\alpha(0)\rangle$ can be expanded in the eigenkets with expansion coefficients $c_m(0)$,

$$|\alpha(0)\rangle = \sum_m c_m(0)|m\rangle, \quad \left(\sum_m |c_m(0)|^2 = 1\right). \tag{9}$$

We wish to find the state ket $|\alpha(t)\rangle$ at any later time $t > 0$ such that

$$|\alpha(t)\rangle = \sum_m c_m(t) e^{-iE_m t/\hbar} |m\rangle, \quad \left(\sum_m |c_m(t)|^2 = 1\right), \tag{10}$$

which is governed by the Schrödinger equation,

$$i\hbar \frac{d}{dt}|\alpha(t)\rangle = (H_0 + h(t))|\alpha(t)\rangle. \tag{11}$$

Substituting (10) into it and multiplying an eigen-bra $\langle n|$ from the left, we have another form of the Schrödinger equation,

$$i\hbar \frac{d}{dt} c_n(t) = \sum_m \langle n|h|m\rangle e^{i\omega_{nm}t} c_m(t) = \sum_m h_{nm} e^{i\omega_{nm}t} c_m(t), \tag{12}$$

where,

$$\omega_{nm} \equiv (E_n - E_m)/\hbar, \quad h_{nm} = \langle n|h|m\rangle. \tag{13}$$

Once the time-dependent expansion coefficients $c_n(t)$ in (10) are obtained, the state $|\alpha(t)\rangle$ of the system at any time is known. However, with the exception of a few problems, exact solutions to the differential equation (12) for $c_n(t)$ are not available.

The perturbation theory offers a ritual way to obtain the approximate solution to Eq.(12) in terms of orders $O(h^r)$, ($r = 0,1,2,3,...$ ) with initial conditions $c_n(0) = c_n^{(0)}(t) \equiv c_n^{(0)}(0)$, and $c_n^{(r+1)}(0) = 0$,,

$$c_n(t) = c_n^{(0)}(t) + c_n^{(1)}(t) + c_n^{(2)}(t) + .... \tag{14}$$

Each higher order solution at later time $c_n^{(r+1)}(t)$ satisfies the following iterative equation,

$$i\hbar \frac{d}{dt} c_n^{(r+1)}(t) = \sum_m h_{nm} e^{i\omega_{nm}t} c_m^{(r)}(t). \tag{15}$$

These equations can in principle be integrated successively to any given order of the perturbative strength, but only the first few orders are practically possible. The formal solutions are given by,

$$c_n^{(r+1)}(t) = \frac{1}{i\hbar} \sum_m \int_0^t h_{nm} e^{i\omega_{nm}t'} c_m^{(r)}(t') dt'. \tag{16}$$

In present paper, we confine ourselves to the system that is initially in a particular unperturbed state $c_n^{(0)}(0) = \delta_{nk}$. We have form Eq. (14),

$$c_n(t) = \delta_{nk} + c_n^{(1)}(t) + c_n^{(2)}(t) + ...$$
$$= \begin{cases} 1 + c_k^{(1)}(t) + c_k^{(2)}(t) + c_k^{(3)}(t) + ... & n = k \\ c_n^{(1)}(t) + c_n^{(2)}(t) + c_n^{(3)}(t) + ... & n \neq k \end{cases}. \tag{17}$$

For a degenerate system, states $|n\rangle$ and $|k\rangle$ with $n \neq k$ label degenerate states of the same energy. The normalization condition requires,

$$1 = \sum_n |c_n(t)|^2 = \left|1 + c_k^{(1)}(t) + c_k^{(2)}(t) + ...\right|^2 + \sum_{n \neq k} \left|c_n^{(1)}(t) + c_n^{(2)}(t) + ...\right|^2. \tag{18}$$

Once the TDPT give all order corrections, the solution converges to the exact solutions without divergence. [6,7] Instead, the approximate solution including first few orders for some systems predicts $|c_n(t)|^2 \gg 1$ as the perturbation is applied for long enough time, which is the secular divergence in quantum mechanics.

## II. 2 Transformation of the perturbative series into complex phases

The relation (18) indicates that both $\left|1+c_k^{(1)}(t)+c_k^{(2)}(t)+...\right|\leq 1$ and $\left|c_n^{(1)}(t)+c_n^{(2)}(t)+...\right|\leq 1$ must be simultaneously satisfied, implying that we can make following transformations. For $n=k$, the first PS $1+\sum_r c_k^{(r)}(t)$ can be transformed into a natural function of a series also in powers of the perturbative strength,

$$1+c_k^{(1)}(t)+c_k^{(2)}(t)+c_k^{(3)}(t)+... = \exp\left(\alpha_k^{(1)}(t)+\alpha_k^{(2)}(t)+\alpha_k^{(3)}(t)+...\right). \quad (19)$$

Similarly, for $n \neq k$, the second PS $\sum_r c_n^{(r)}(t)$ is,

$$c_n^{(1)}(t)+c_n^{(2)}(t)+c_n^{(3)}(t)+...=(\mathrm{Re}+i\,\mathrm{Im})\left(c_n^{(1)}(t)+c_n^{(2)}(t)+c_n^{(3)}(t)+...\right)$$

$$\begin{aligned}&= \sin\left[\mathrm{Re}\left(\beta_n^{(1)}(t)+\beta_n^{(2)}(t)+\beta_n^{(3)}(t)+...\right)\right]\\&+i\sin\left[\mathrm{Im}\left(\beta_n^{(1)}(t)+\beta_n^{(2)}(t)+\beta_n^{(3)}(t)+...\right)\right],\end{aligned} \quad (20)$$

where

$$\mathrm{Pr}\left(\sum_{r=1}\beta_n^{(r)}(t)\right) = \arcsin\left(\mathrm{Pr}\left(\sum_{r=1}c_n^{(r)}(t)\right)\right),\ (\mathrm{Pr}=\mathrm{Re},\mathrm{Im}). \quad (21)$$

The first three terms in $\sum_r \alpha_k^{(r)}(t)$ are,

$$\alpha_k^{(1)}(t)=c_k^{(1)}(t),\ \alpha_k^{(2)}(t)=c_k^{(2)}(t)-\frac{c_k^{(1)}(t)^2}{2},$$
$$\alpha_k^{(3)}(t)=\frac{c_k^{(1)}(t)^3}{3}-c_k^{(1)}(t)c_k^{(2)}(t)+c_k^{(3)}(t),\ ... \quad (22)$$

And those in $\sum_n \beta_k^{(n)}(t)$ are,

$$\beta_n^{(1)}(t)=c_n^{(1)}(t),\ \beta_n^{(2)}(t)=c_n^{(2)}(t),\ \beta_n^{(3)}(t)=c_n^{(3)}(t)+\frac{c_n^{(1)}(t)^3}{6},\ .... \quad (23)$$

In sum, instead of using Eq. (17) as the PS, I propose to use following functions of series,

$$c_n(t) = \begin{cases} \exp\left(\sum_k \alpha_k^{(r)}(t)\right), & n \neq k \\ \sin\left\{\operatorname{Re}\sum_r \beta_n^{(r)}(t)\right\} + i\sin\left\{\operatorname{Im}\sum_r \beta_n^{(r)}(t)\right\}, & n \neq k \end{cases}, \quad (24)$$

where $\alpha_k^{(r)}(t)$ and $\beta_n^{(r)}(t)$ are complex functions. To make this new form of perturbation solution evidently self-consistent, we can directly substitute the left-hand side of Eqs. (24) into the Eq.(12), and establish the iterative equations for $\alpha_k^{(r)}(t)$ and $\beta_k^{(r)}(t)$ that are identical to those transformed from the original PS via (19) and (20). The resultant probabilities are bounded from above as time approaches to infinity.

### III, Three exactly solvable models

#### III. 1 A two level system with constant perturbation

Two parts of the Hamiltonian $H = H_0 + h(t)$ are, [8]

$$H_0 = \begin{pmatrix} E_1 & 0 \\ 0 & E_2 \end{pmatrix}, \quad (0 < E_1 < E_2), \quad h = \begin{pmatrix} 0 & \gamma \\ \gamma & 0 \end{pmatrix} \quad (\gamma \ll E_1), \quad (t > 0). \quad (25)$$

Initially, the system is in the ground state $\varphi_1 = (1, 0)^T$, and we hope to examine the probability of the system in this state after sufficiently long period the perturbation switches on.

We find from (16) and (17),

$$c_1^{(1)} = 0, \quad c_1^{(2)} = i\frac{\gamma^2}{\hbar^2\omega^2}(\omega t - \sin\omega t) - \frac{2\gamma^2}{\hbar^2\omega^2}\sin^2\frac{\omega t}{2}, \quad (26)$$

where $\omega \equiv (E_2 - E_1)/\hbar$. We see that in $c_1^{(2)}$ there is an imaginary term linear in time. The probability amplitude $c_1(t)$ of second order accuracy is,

$$c_1(t) \approx 1 + c_1^{(1)} + c_1^{(2)} = 1 + \left( i\frac{\gamma^2}{\hbar^2\omega^2}(\omega t - \sin\omega t) - \frac{2\gamma^2}{\hbar^2\omega^2}\sin^2\frac{\omega t}{2} \right) \quad (27)$$

Then, the probability increases as $t^2$ for a long enough time $t$,

$$p_1 \approx \left|1 + c_1^{(1)} + c_1^{(2)}\right|^2 = \left(1 - \frac{2\gamma^2}{\hbar^2\omega^2}\sin^2\frac{\omega t}{2}\right)^2 + \left(\frac{\gamma^2}{\hbar^2\omega^2}(\omega t - \sin\omega t)\right)^2 \to t^2. \quad (28)$$

However, this absurd increase can be easily eliminated by use of my formulation (19),

$$\begin{aligned}
c_1(t) &\approx \exp\left[c_1^{(1)} + \left(c_1^{(2)} - \left(c_1^{(1)}\right)^2/2\right)\right] = \exp\left(c_1^{(2)}\right) \\
&= \exp\left( i\frac{\gamma^2}{\hbar^2\omega^2}(\omega t - \sin\omega t) - \frac{2\gamma^2}{\hbar^2\omega^2}\sin^2\frac{\omega t}{2} \right) \\
&\approx \exp\left( i\frac{\gamma^2}{\hbar^2\omega^2}\omega t \right) - i\frac{\gamma^2}{\hbar^2\omega^2}\sin\omega t - \frac{2\gamma^2}{\hbar^2\omega^2}\sin^2\frac{\omega t}{2} \\
&= \exp\left( i\frac{\gamma^2}{\hbar^2\omega^2}\omega t \right) - \frac{\gamma^2}{\hbar^2\omega^2}\left(1 - e^{-i\omega t}\right).
\end{aligned} \quad (29)$$

Because the problem (25) is exactly solvable, and the solution is, [8]

$$\psi(t) = \left\{\left(\cos\frac{\Omega t}{2} + i\frac{\omega}{\Omega}\sin\frac{\Omega t}{2}\right)\varphi_1 + \left(-i\frac{2\gamma}{\Omega}\right)\sin\frac{\Omega t}{2}\varphi_2\right\}\exp\left\{-\frac{i}{2}\frac{(E_1+E_2)t}{\hbar}\right\} \quad (30)$$

where $\Omega \equiv \sqrt{\omega^2 + 4\gamma^2/\hbar^2}$, and $\varphi_2 = (0,1)^T$. The approximate solution up to second order of $\gamma^2$ for the system in the ground state $\varphi_1 = (1,0)^T$ is given by,

$$\exp\left(-\frac{iE_1 t}{\hbar}\right)\left\{\exp\left(\frac{i\gamma^2}{\hbar^2\omega^2}\omega t\right) - \frac{\gamma^2}{\hbar^2\omega^2}\left(1 - e^{-i\omega t}\right)\right\}. \quad (31)$$

It is exactly the approximate result accurate to the second order of $\gamma$ (29), with noting the phase factor $e^{-E_n t/\hbar}$ being left before $|n\rangle$ in Eq. (10).

### III. 2 A two level system with constant perturbation: degenerate case

Two parts of the Hamiltonian $H = H_0 + h(t)$ are, [9]

$$H_0 = \begin{pmatrix} E_0 & 0 \\ 0 & E_0 \end{pmatrix}, \quad h = \begin{pmatrix} 0 & \gamma \\ \gamma & 0 \end{pmatrix} \ (\gamma \ll E_0), \ (t > 0). \tag{32}$$

Initially, the system is in state $\varphi_1 = (1, 0)^T$ which is doubly degenerate. We need to examine the probability of the system in another degenerate state $\varphi_2 = (0, 1)^T$ after sufficiently long time of the perturbation acted.

The first order correction $c_2^{(1)}(t)$ is of imaginarily linear dependence on time

$$c_2^{(1)}(t) = \frac{\gamma}{i\hbar} t. \tag{33}$$

It seems to give the divergent transition probability,

$$\left| c_2^{(1)}(t) \right|^2 = \frac{\gamma^2}{\hbar^2} t^2, \tag{34}$$

whereas our Eq. (20) gives,

$$c_n^{(1)}(t) = \frac{1}{i} \sin(\frac{\gamma}{\hbar} t), \tag{35}$$

which agrees with Eq. (34) only for short periods of time such as $|\gamma t / \hbar| \approx 0$. Result (35) is the same as that given by the exact solution that is discussed in detail by Bransden and Joachim, [9] and Feynman and Hibbs. [6] However, Bransden and Joachim conclude that the result (34) indicates "that perturbation theory is breaking down". [9] From the point of my formulation, this conclusion is true but superficial.

Multiplying the phase factor $e^{-iE_0 t/\hbar}$ and $c_n^{(1)}(t)$, we have $c_n^{(1)}(t) e^{-iE_0 t/\hbar}$ $\sim e^{-i(E_0 - \gamma)t/\hbar} - e^{-i(E_0 + \gamma)t/\hbar}$. Thus, it is not an overall renormalization of the energy eigenvalues.

### III. 3  Spin resonance

Two parts of the Hamiltonian $H = H_0 + h(t)$ are, [2]

$$H_0 = -\frac{\hbar}{2}\eta B_0 \sigma_z, \quad h = -\frac{\hbar}{2}\eta B_1\left(\sigma_x \cos(\omega t) + \sigma_y \sin(\omega t)\right) \quad (B_1 \ll B_0), \quad (36)$$

where $\sigma_i$, $(i=1,2,3)$ are Pauli matrices, and $B_j$, $(j=0,1)$ stand for two components of the magnetic field, and $\eta > 0$ is a positive constant. The initial state at $t=0$ is chosen to be spin-up $\varphi_{up} = (1,0)^T$ and the energy is $\varepsilon_i = -\hbar\eta B_0/2$, and we now explore that at the end of the interval $0 \le t \le T$ during the perturbation acts, what the probability amplitude is of the system in spin-down $\varphi_{down} = (0,1)^T$, whose energy is $\varepsilon_f = \hbar\eta B_0/2$.

The first order correction $c_2^{(1)}(t)$ is after calculations, [2]

$$c_f^{(1)}(t) = \frac{1}{i\hbar}\int_0^t \langle f|h|i\rangle e^{i\omega_0 t'} dt' = \frac{1}{i}\int_0^T \left(-\frac{\omega_1}{2}\right) e^{i(\omega_0+\omega)t'} dt' \\ = -\frac{\omega_1}{2}\frac{1-\exp[i(\omega_0+\omega)T]}{\omega_0+\omega} \quad (37)$$

where $\omega_0 = \eta B_0$ and $\omega_1 = \eta B_1$. At resonance $\omega_0 + \omega = 0$, we have,

$$c_f^{(1)}(t) = \frac{1}{i}\int_0^T \left(-\frac{\omega_1}{2}\right) e^{i(\omega_0+\omega)t'} dt' = i\frac{\omega_1}{2}T \quad (38)$$

We see again the imaginarily linear dependence on time. The directly squared norm of result (38) gives,

$$\left|c_n^{(1)}(t)\right|^2 = \left(\frac{\omega_1 T}{2}\right)^2, \quad (39)$$

which is usually interpreted as the transition probability, but holds true only for short periods of time. Our Eq. (20) gives the correct result,

$$c_n^{(1)}(t) = i\sin(\frac{\omega_1}{2}T), \text{ and } \left|c_n^{(1)}(t)\right|^2 = \left\{\sin\left(\frac{\omega_1}{2}T\right)\right\}^2. \quad (40)$$

This is also the same as that given by the exact solution.[2]

## IV. Conclusions and discussions

To remove the secular divergences associated with the conventional form of the TDPT in quantum mechanics, a transformation of the naïve form of PS into natural exponential functions of another PS is proposed. Three exactly solvable models are used to verify the new formulation which gives results that are excellently agreement with the exact ones.

It appears very curious for such a simple replacement of the naïve form of PS by its *equivalent* natural exponential functions of another PS is so effective and powerful. In fact, the current form of the TDPT is more a physics method than a mathematics theory.[3] Only in the following limited sense the naïve form of PS and the transformed form of the PS as complex phases are *equivalent*: Both expansions are in powers of perturbative strength rather than the time or the magnitude of the each term in the PS. However, in comparison of the calculated results with experimental data one needs to deal with the probability in which the time plays a crucial role. Therefore we have to pay attention as well on the role of the time in the TDPT from the beginning. To note that in the Schrödinger equation, the time is formally conjugate to the Hamiltonian and always associated with the imaginary number unit $i$. Once the perturbation presents and the PS is taken as complex phases, the action of the perturbation turns out to have finite contributions into the time-dependent probabilities, as expected. This is why the natural exponential functions of PS are more powerful than the naïve form that holds true only for short period of time. However, I do not think either such a simple transformation is universally applicable, because in present paper we only deal with the cases that the system is initially in a particular unperturbed state $c_n^{(0)}(0) = \delta_{nk}$; or such a transformation will be no use in other cases because we do not know in what extents our formulation is still feasible, which is still under explorations.

## Acknowledgments

This work is financially supported by National Natural Science Foundation of China under Grant No. 11175063.

* Author to whom correspondence should be addressed. Electric address: quanhuiliu@gmai.com

[1] W. Pauli, *General Principles of Quantum Mechanics* (Springer-Verlag, Berlin, 1980) p.87.

[2] L. E. Ballentine, *Quantum Mechanics A Modern Development* (World Scientific, Singapore,1998) p.354.

[3] P. W. Langhoff, S. T. Epstein, and M. Karplus, "Aspects of time-dependent perturbation theory", Rev. Mod. Phys., 44(3), 602-644(1972).

[4] K. Bhattacharyya, A Modification of the Dirac Time-Dependent Perturbation Theory, Proc. Royal Soc. London. A, 394(8), 345-361 (1984).

[5] J. K. Bhattacharjee and D. S. Ray, "Time-dependent perturbation theory in quantum mechanics and the renormalization group," Am. J. Phys., 84(6), 434 -442(2016).

[6] R. P. Feynman, A.R. Hibbs，*Quantum Mechanics and Path Integrals* (McGraw-Hill, N. Y., 1965 ), p.149.

[7] C. Cohen-Tannoudji, B. Diu, F. Laloe, *Quantum Mechanics* ,*Vol. 2* (John Wiley，N. Y. 1977), p.1290.

[8] B.C. Qian，J. Y. Zeng, *Problems and solutions in quantum mechanics* (3$^{rd}$ ed.) (Science Press, Beijing, 2008). Ex. 13.4. p.382.

[9] B. H. Bransden and C. J. Joachim, *Introduction to Quantum Mechanics* (Longman, London, 1989), pp.414-426.